\documentclass[12pt]{article}

\usepackage{amssymb}
\usepackage{amsmath}
\usepackage{amsfonts}

\newcommand{\R}{\mathbb{R}}
\newcommand{\C}{\mathbb{C}}

\newcommand{\be}{\begin{equation}}
\newcommand{\bea}{\begin{eqnarray}}
\newcommand{\eea}{\end{eqnarray}}

\newcommand{\kt}{\rangle}
\newcommand{\br}{\langle}

\newcommand{\ed}{\end{document}}
\newcommand{\bbr}{\br\!\br}
\newcommand{\kkt}{\kt\!\kt}
\newcommand{\pbr}{\prec}
\newcommand{\pkt}{\succ}

\begin{document}

\title{Wave Function of the Universe and~Its~Meaning}
\author{\\
Ali Mostafazadeh\thanks{E-mail address: amostafazadeh@ku.edu.tr}\\
\\ Department of Mathematics, Ko\c{c} University,\\
Rumelifeneri Yolu, 34450 Sariyer,\\
Istanbul, Turkey}
\date{ }
\maketitle

\begin{abstract}
For a FRW-spacetime coupled to an arbitrary real scalar field, we
endow the solution space of the associated Wheeler-DeWitt equation
with a Hilbert-space structure, construct the observables, and
introduce the physical wave functions of the universe that admit a
genuine probabilistic interpretation. We also discuss a proposal
for the formulation of the dynamics. The approach to quantum
cosmology outlined in this article is based on the results
obtained within the theory of pseudo-Hermitian operators.
\end{abstract}

\section{Introduction}

Quantum cosmology is a natural outcome of the efforts toward a
unification of quantum mechanics (QM) and general relativity,
i.e., development of a quantum theory of gravity (QG)
\cite{dewitt,review,wiltshire}. Even in its gravely simplified
minisuperspace realizations \cite{dewitt,misner,review,wiltshire},
quantum cosmology provides a useful testing ground for various
proposals for solving some of the most important problems of QG
such as the Hilbert-space problem, the factor-ordering problem,
and the problem of time. These problems have been studied
extensively for the past four decades. Yet, even for the simplest
minisuperspace models, their complete solution could not be found.

This article aims at providing a brief outline of a consistent
formulation of quantum cosmology, based on a FRW-spacetime coupled
to an arbitrary real scalar field, that offers explicit solutions
for some of these problems and provides valuable insight in
others.

\section{Quantization of the model and the Wheeler-DeWitt equation}

Consider a FRW-spacetime coupled to a real scalar field $\varphi$
with an arbitrary real-valued potential $V=V(\varphi)$. The
Einstein equations for this model are equivalent to a single
differential equation which can be cast in the form of a
Hamiltonian constraint \cite{isham-75}. Choosing the natural
system of units described in \cite{wiltshire}, the latter takes
the form
    \be
    {\cal K}:=-\pi_\alpha^2+\pi_\varphi^2-\kappa\,e^{4\alpha}+
    e^{6\alpha}V(\varphi)=0,
    \label{constraint}
    \end{equation}
where $\alpha:=\ln a$, $a$ is the scale factor of the FRW model,
$\kappa$ is the curvature index with the values $-1,0,1$ that
respectively correspond to an open, flat, or closed universe, and
$\pi_\alpha$ and $\pi_\varphi$ are the canonical momenta conjugate
to $\alpha$ and $\varphi$.

The standard canonical quantization of the above model uses
Dirac's method of quantizing constrained systems \cite{dirac}.
This involves the canonical quantization of the unconstrained
system and the imposition of the constraint as a restriction on
the allowed state vectors. For the system described by the
phase-space variables $(\alpha,\pi_\alpha;\varphi,\pi_\varphi)$
and the constraint (\ref{constraint}), this leads to the auxiliary
Hilbert space ${\cal H}'=L^2(\R^2)$ and the operators
$(\hat\alpha',\hat\pi_\alpha';\hat\varphi',\hat\pi_\varphi')$ that
act in ${\cal H}'$ and satisfy the canonical commutation relations
$[\hat\alpha',\hat\varphi']=[\hat\pi_\alpha',\hat\pi_\varphi']=
[\hat\alpha',\hat\pi_\varphi']=[\hat\varphi',\hat\pi_\alpha']=0$
and $[\hat\alpha',\hat\pi_\alpha']=
[\hat\varphi',\hat\pi_\varphi']=i$. Moreover, the quantum analogue
of the classical constraint~(\ref{constraint}) takes the form
$\hat{\cal K}'|\psi)=0$, where $\hat{\cal K}'$ is obtained by
quantizing the classical Hamiltonian ${\cal K}$ and has, up to the
factor-ordering ambiguities, the form: $\hat{\cal
K}'=-\hat{\pi'}_\alpha^2+\hat{\pi'}_\varphi^2+e^{6\hat\alpha'}
V(\hat\varphi')-\kappa~e^{4\hat\alpha'}$.

A few remarks are in order: 1.~The quantum constraint, $\hat{\cal
K}'|\psi)=0$, identifies the space of the physical state vectors
of the system with the kernel ${\cal V}$ of the operator
$\hat{\cal K}'$. In particular, ${\cal V}$ is a vector subspace of
the auxiliary Hilbert space $L^2(\R^2)$. Because Dirac's
quantization scheme does not endow ${\cal V}$ with an inner
product, one must find a way to construct an inner product on
${\cal V}$ and promote it to a genuine Hilbert space ${\cal H}$.
This is known as the {\em Hilbert-space problem}. 2.~The
observables of the theory are the Hermitian operators acting in
${\cal H}$; 3.~The usual formulation of quantum dynamics requires
identifying one of the observables with the Hamiltonian operator
and defining the evolution parameter $\tau$ appearing in the
corresponding Schr\"odinger equation with time.

It is usually more instructive to write the quantum constraint as
a differential equation. To do this, one first identifies $\vec
x':=(\hat\alpha',\hat\varphi')$ with the position operator acting
in ${\cal H}'$ and expresses the state vectors $|\psi)$ and the
relevant operators in the position representation. The
($\delta$-function normalized) position basis kets
$|\alpha',\varphi')$ satisfy the defining relations:
$(\alpha',\varphi'|\hat\alpha'=\alpha'(\alpha',\varphi'|$,
$(\alpha',\varphi'|\hat\varphi'=\varphi'(\alpha',\varphi'|$,
$(\alpha',\varphi'|\hat\pi'_\alpha=
-i\partial_{\alpha'}(\alpha',\varphi'|$,
$(\alpha',\varphi'|\hat\pi'_\varphi=
-i\partial_{\varphi'}(\alpha',\varphi'|$. In this basis, the
elements $|\psi)$ of ${\cal H}'$ are represented by the `position'
wave functions
    \be
    \psi(\alpha',\varphi'):=(\alpha',\varphi'|\psi),
    \label{wf-1}
    \end{equation}
and the quantum constraint, $\hat{\cal K}'|\psi)=0$, takes the
form of the Wheeler-DeWitt equation
\cite{dewitt,review,wiltshire}:
    \be
    \left[-\partial_{\alpha'}^2+\partial_{\varphi'}^2+
    \kappa\,e^{4\alpha'}
    -e^{6\alpha'}V(\varphi')\right]\psi(\alpha',\varphi')=0.
    \label{wdw}
    \end{equation}
This in turn implies that one can identify ${\cal V}$ with the
(vector) space of solutions of this equation. Indeed, as far as
the Dirac's program of constrained quantization is concerned, the
function of the Wheeler-DeWitt equation is to determine ${\cal
V}$.

\section{Hilbert-space problem}

Because (\ref{wdw}) is a second order hyperbolic equation, its
solutions $\psi\in{\cal V}$ may be uniquely determined in terms of
two initial conditions. This is done by selecting a time-like
coordinate $\tau'$ for the $(1+1)$-dimensional Minkowski space
parameterized by the coordinates $(\alpha',\varphi')$ and
specifying each solution $\psi$ with a pair of initial data for
(\ref{wdw}) at some initial value $\tau_0'$ of $\tau'$. The choice
of the $\tau'$ and $\tau_0'$ is actually arbitrary. Here we will
take $\tau':=\alpha'$, but we should like to emphasize that this
does not mean that we identify $\alpha'$ with a physical time
variable. We will determine the latter by requiring that it is an
evolution parameter for a Hamiltonian operator acting in the
physical Hilbert space ${\cal H}$.

Next, let $\alpha'_0\in\R$ be an arbitrary initial value for
$\alpha'$, and for all $\psi\in{\cal V}$ and $\alpha'\in\R$ define
$\psi(\alpha'):\R\to\C$ according to $\psi(\alpha')[\varphi']:=
\psi(\alpha',\varphi')$, where $\varphi'\in\R$ is arbitrary. Then
we can view both $\psi(\alpha')$ and
$\dot\psi(\alpha'):=\partial_{\alpha'}\psi(\alpha')$ as elements
of $L^2(\R)$. In particular, we have
$\psi(\alpha_0'),\dot\psi(\alpha_0')\in L^2(\R)$. Furthermore, we
can express the Wheeler-DeWitt equation~(\ref{wdw}), in the form
    \be
    [\partial_{\alpha'}^2+D]\psi(\alpha')=0,
    \label{kgt}
    \end{equation}
where $D:L^2(\R)\to L^2(\R)$ is the Hermitian operator defined by
$(D\xi)(\varphi'):= [-\partial_{\varphi'}^2+e^{6\alpha'}
V(\varphi')-\kappa\,e^{4\alpha'}]\xi(\varphi')$, with $\xi\in
L^2(\R)$ and $\varphi'\in\R$.

Having identified the Wheeler-DeWitt equation~(\ref{wdw}) with the
second order equation (\ref{kgt}), which is defined in the Hilbert
space $L^2(\R)$, we can write ${\cal V}=\{\psi:\R\to
L^2(\R)|[\partial_{\alpha'}^2+D]\psi(\alpha')=0\}$ and use the
initial data: $(\psi(\alpha_0),\dot\psi(\alpha_0))\in
L^2(\R)\times L^2(\R)$ to specify the elements $\psi$ of ${\cal
V}$. Because (\ref{kgt}) is a linear equation, as a complex vector
space, ${\cal V}$ is isomorphic to $L^2(\R)\oplus L^2(\R)$. On the
other hand, endowing ${\cal V}$ with any inner product so that it
acquires a separable Hilbert-space structure yields a Hilbert
space ${\cal H}$ that is also isomorphic to $L^2(\R)\oplus
L^2(\R)$. This follows from the uniqueness of the structure of
separable Hilbert spaces \cite{reed-simon}. This observation
reduces the Hilbert-space problem for the model considered here to
the construction of a positive-definite inner product on ${\cal
V}$. An explicit example is \cite{cqg,p54}
    \be
    \pbr\psi_1,\psi_2\pkt:=\frac{1}{2}\left[\br\psi_1(\alpha_0'),D_0
    \psi_2(\alpha_0')\kt+\br\dot\psi_1(\alpha_0'),
    \dot\psi_2(\alpha_0')\kt\right],
    \label{inn}
    \end{equation}
where $\br\cdot,\cdot\kt$ is the inner product of $L^2(\R)$ and
$D_0$ is an arbitrary positive-definite operator (a Hermitian
operator with a strictly positive spectrum) acting in $L^2(\R)$.
One can check that (\ref{inn}) is a positive-definite inner
product on ${\cal V}$ and that any other inner product on this
space is unitarily equivalent to (\ref{inn}), \cite{p54}. The
Hilbert space ${\cal H}$, which consists of the physical state
vectors of the model and is called the physical Hilbert space, is
the Cauchy completion of the inner product space obtained by
endowing ${\cal V}$ with the inner product (\ref{inn}).

It must be emphasized that although the inner product (\ref{inn})
depends on $\alpha_0'$ and $D_0$, the structure of the Hilbert
space ${\cal H}$ is insensitive to the choice of $\alpha_0'$ and
$D_0$; different choices yield unitarily equivalent Hilbert spaces
and consequently the same physical theory.

We conclude this section by noting that the original construction
of the inner product (\ref{inn}) and its
generalizations~\cite{cqg} have their root in the results obtained
in the context of a recently developed theory of pseudo-Hermitian
operators \cite{p1-p7}.

\section{Observables and the wave functions of the universe}

The unitary-equivalence of the physical Hilbert space ${\cal H}$
and $L^2(\R)\oplus L^2(\R)$ means that there exists a linear
(invertible) operator $U:{\cal H}\to L^2(\R)\oplus L^2(\R)$ such
that for all $\psi_1,\psi_2\in{\cal H}$, $\pbr\psi_1,\psi_2\pkt=
\br U\psi_1|U\psi_2\kt$, where $\br\cdot|\cdot\kt$ stands for the
inner product of $L^2(\R)\oplus L^2(\R)$. The operator $U$ is
clearly not unique. Explicit form of such an operator is given in
\cite{p54}. But as we shall see below the form of this operator
does not have a physical significance.

One may use $U$ to relate the Hermitian operators acting in ${\cal
H}$ to those acting in $L^2(\R)\oplus L^2(\R)$. This means that
the physical observables $\hat O$ of the quantum cosmological
model under study have the form $\hat O=U^{-1}\hat o U$ where
$\hat o$ is a Hermitian operator acting in $L^2(\R)\oplus
L^2(\R)$. For instance, we have the basic observables: $\hat
Q:=U^{-1}\hat q\, U$, $\hat P=U^{-1}\hat p\, U$, and $\hat
S_\mu:=U^{-1}\hat s_\mu U$, where $\mu=0,1,2,3$,
$s_\mu:=\sigma_\mu\otimes\hat 1$, $\sigma_0$ is the $2\times 2$
identity matrix, $\sigma_\mu$ with $\mu\neq 0$ are Pauli matrices,
and $\hat 1$, $\hat q$ and $\hat p$ are respectively the identity,
position, and momentum operators acting in $L^2(\R)$. In
particular, $\hat Q$ and $\hat S_3$ form a maximal set of
commuting observables. Hence the corresponding eigenvectors
$\psi^{(q,\nu)}$, with $q\in\R$ and $\nu=\pm 1$, form a `basis' of
${\cal H}$. Clearly $\psi^{(q,\nu)}=U^{-1}|q,\nu\kt$, where
$|q,\nu\kt$ satisfy the defining relations: $\hat
q|q,\nu\kt=q|q,\nu\kt$ and $\hat s_3|q,\nu\kt=\nu |q,\nu\kt$, and
the orthonormality and completeness relations $\br
q,\nu|q',\nu'\kt= \delta(q-q')\delta_{\nu,\nu'}$ and
$\sum_{\nu=\pm 1}\int_{-\infty}^\infty dq\,|q,\nu\kt \br
q,\nu|=\hat s_0$, respectively.

We can express any element $\psi$ of ${\cal H}$ in the basis
$\{\psi^{(q,\nu)}\}$ as
    \be
    \psi=\sum_{\nu=\pm 1}\int_{-\infty}^\infty
    dq\,\Psi(q,\nu)\:\psi^{(q,\nu)},
    \label{wf}
    \end{equation}
where $\Psi:\R\times\{-1,1\}\to\C$ are the coefficient functions.
One can easily show that indeed $\Psi$ belong to $\in
L^2(\R\times\{-1,1\})$ which is isomorphic, as a Hilbert space, to
$L^2(\R)\oplus L^2(\R)$.

The arguments $q$ and $\nu$ of $\Psi$ are the eigenvalues of the
physical observables $\hat Q$ and $\hat S_3$. This is not true for
the arguments $\alpha'$ and $\varphi'$ of the wave functions
$\psi(\alpha',\varphi')$ appearing in the Wheeler-DeWitt
equation~(\ref{wdw}), because these are the eigenvalues of the
operators $\hat\alpha'$ and $\hat\varphi'$ which act in the
auxiliary Hilbert space ${\cal H}'$ and do not leave the physical
Hilbert space ${\cal H}$ invariant, i.e., there are $\psi\in{\cal
H}$ such that $\hat\alpha'\psi,\hat\varphi'\psi\notin{\cal H}$.
This shows that the description of the state vectors $\psi$ as
functions depending on $(\alpha',\varphi')$ is quite different
from the description of the state vectors in ordinary QM in terms
of the position wave functions. It is the coefficient functions
$\Psi$ that have the eigenvalues of physical observables as their
argument and play the role of the familiar position wave
functions. Therefore, we propose to refer to them as the physical
``Wave Function of the Universe.''

The expression~(\ref{wf}) provides a one-to-one correspondence
between the state vectors $\psi$ and the wave functions $\Psi$.
This is a manifestation of the unitary-equivalence of ${\cal H}$
and $L^2(\R)\oplus L^2(\R)$. Indeed one may use (\ref{wf}) to
define ${\cal U}:{\cal H}\to L^2(\R)\oplus L^2(\R)$ according to
${\cal U}\psi:=\mbox{\footnotesize$
\left(\begin{array}{c}\Psi(q,+)\\
\Psi(q,-)\end{array}\right)$}$ with $\Psi(q,\nu):=
\bbr\psi^{(q,\nu)},\psi\kkt$, and check that ${\cal U}$ is a
unitary operator. An immediate implication of the existence and
unitarity of ${\cal U}$ is that, similarly to ordinary QM, we can
formulate our quantum cosmological theory in terms of the wave
functions ${\cal U}\psi$ which belong to the well-known Hilbert
space $L^2(\R)\oplus L^2(\R)$. We can express the inner product of
two state vectors $\psi_1,\psi_2\in{\cal H}$ in terms of their
wave functions $\Psi_1$ and $\Psi_2$ as
$\pbr\psi_1,\psi_2\pkt=\sum_{\nu=\pm 1} \int_{-\infty}^\infty dq\:
\Psi_1(q,\nu)^*\Psi_2(q,\nu)$. Similarly, we can express the
action of an observable $\hat O$ on a state vector $\psi\in{\cal
H}$ in terms of the corresponding wave function $\Psi$ by defining
$\hat\Omega:={\cal U}\hat O\,{\cal U}^{-1}:L^2(\R)\oplus
L^2(\R)\to L^2(\R)\oplus L^2(\R)$ and checking that indeed $\hat
O\psi=\sum_{\nu=\pm 1} \int_{-\infty}^\infty
dq\,[\hat\Omega\Psi(q,\nu)] \:\psi^{(q,\nu)}$. For example, we
have $\hat Q\psi=\sum_{\nu=\pm 1} \int_{-\infty}^\infty
dq\,q\Psi(q,\nu)]\:\psi^{(q,\nu)}$. The converse of the above
argument also holds: Every observable $\hat\Omega$ acting in
$L^2(\R)\oplus L^2(\R)$ defines an observable $\hat O:={\cal
U}^{-1}\hat\Omega\,{\cal U}$ acting in ${\cal H}$.

\section{Physical Interpretation and the Dynamics}

So far we have constructed the Hilbert space and obtained the form
of the observables, but failed to describe how the latter connect
with the classical observables or discuss their physical meaning.
Another crucial issue that we have not treated is the dynamics.

We propose to associate physical meaning to the basic observables
$\hat Q$ and $\hat S_3$ and consequently the variables $q$ and
$\nu$ by identifying the former with quantum analogues of a pair
of observables for the reduced (constrained) classical system
($C$). This means that we intend to perform a quantization of $C$
that leads to a quantum theory having ${\cal H}$ as its Hilbert
space. Similarly, we propose to determine a Hamiltonian operator
by selecting a classical time variable, obtaining a classical
Hamiltonian that describes the dynamics of $C$, and carrying out
the above-mentioned quantization.

A widely used classical time is the logarithm of the scale factor,
i.e., $\alpha$, which is acceptable if the classical universe is
ever-expanding. There are classical universes that violate this
assumption. It is also obvious that one cannot make such an
assumption in a quantum mechanical treatment of the universe. An
alternative to taking $\alpha$ as a classical time-variable that
avoids this assumption is $\tau:=\epsilon\alpha$, where
$\epsilon:=(d\alpha(t)/dt)/|d\alpha(t)/dt|$. Clearly $\tau$ is a
monotonically increasing function of the physical time $t$.

If we choose $\tau$ as a classical time-variable, we can generate
the dynamics of $C$ using the classical Hamiltonian
$H=\sqrt{\pi_\varphi^2+e^{6\epsilon\tau}V(\varphi)-\kappa\,
e^{4\epsilon\tau}}$. As seen from this relation $C$ has $\varphi$
and $\epsilon$ as configuration variables. This suggests the
following quantization rule: $\varphi\to\hat Q$,
$\pi_\varphi\to\hat P$, and $\epsilon\to\hat S_3$. For
convenience, we introduce: $\hat\varphi:=\hat Q$,
$\hat\pi_\varphi:=\hat P$, and $\hat\epsilon:=\hat S_3$, which in
turn suggest the following identifications: $q=\varphi$ and
$\nu=\epsilon$.

Aside from the usual factor-ordering ambiguities that are also
present in ordinary QM, the above quantization rule leads to the
Hamiltonian operator
    \be
    \hat H=\sqrt{\hat\pi_\varphi^2+e^{6\hat\epsilon\tau}V(\hat\varphi)
    -\kappa\,e^{4\hat\epsilon\tau}}.
    \label{H=}
    \end{equation}
The dynamics of the quantum universe is then determined by the
Schr\"odinger equation $id\psi_\tau/d\tau=\hat H\psi_\tau$, where
we have expressed the time-dependence of the evolving state vector
$\psi_\tau$ using the index $\tau$. We can also express the
dynamics in terms of the physical wave functions $\Psi$ whose
arguments $q$ and $\nu$ are respectively identified with the
scalar field $\varphi$ and the expansion index $\epsilon$. Letting
$\Psi(\varphi,\epsilon;\tau):=
\pbr~\psi^{(\varphi,\epsilon)},\psi_\tau\pkt$, we have \cite{p54}
    \be
    i\partial_\tau\,\Psi(\varphi,\epsilon;\epsilon\tau)=
    \epsilon\sqrt{-\partial_\varphi^2+ e^{6\tau}V(\varphi)-
    \kappa\,e^{4\tau}}\;\Psi(\varphi,\epsilon;\epsilon\tau).
    \label{sch-eq-wf}
    \end{equation}

As seen from (\ref{H=}), the quantum theory described by the
Hilbert space ${\cal H}$ and the Hamiltonian $\hat H$ is a unitary
theory admitting a probabilistic interpretation, if the operator
appearing in the square root in (\ref{H=}) (alternatively in
(\ref{sch-eq-wf})) is a positive operator. This is always the case
for the class of open and flat FRW models coupled to a real scalar
field with a positive confining potential, i.e., whenever
$\kappa\neq 1$, $V(\varphi)\geq 0$ for all $\varphi\in\R$, and
$\lim_{|\varphi|\to\infty}V(\varphi)=\infty$. Various models that
allow for classical inflationary expansions belong to this class.
Typical examples are open and flat FRW models coupled to a massive
real scalar field \cite{wiltshire}.

In general $\hat H$ fails to be a Hermitian operator for a range
of values of the time-variable. Outside this range the quantum
theory is unitary. The description of the physics of crossing the
boundary of this range is still open to both quantitative and
qualitative investigation.

\section{Conclusion}
In this article we presented a summary of our recent attempts to
overcome some of the most fundamental problem of quantum cosmology
for a large family of cosmological models. Perhaps the most
important feature of our method is its formulation of the
Wheeler-DeWitt equation as a second order ordinary differential
equation defined in a Hilbert space $\tilde{\cal H}$. This is
equivalent to a first order equation defined in $\tilde{\cal
H}\oplus\tilde{\cal H}$ which may be identified with a
Schr\"odinger equation with a generally non-Hermitian but
pseudo-Hermitian Hamiltonian. It is the basic spectral properties
of the pseudo-Hermitian Hamiltonians \cite{p1-p7} that lead to the
results reported here.

The following are our concluding remarks: 1.~The construction of
the Hilbert space and subsequently the observables is insensitive
to the particular factor-ordering prescription chosen to write
down the Wheeler-DeWitt equation~(\ref{wdw}). The factor-ordering
problem only arises while quantizing the Hamiltonian of the
(reduced) classical system. It is effective to the same extend as
in the ordinary QM; 2.~A notable feature of our investigation is
the introduction of the physical wave functions
$\Psi(\varphi,\pm;\tau)$ of the universe that effectively describe
the expanding and retracting components of the quantum state of
the universe. The whole theory may be described in terms of these
wave functions; 3.~Our approach may be viewed as providing a link
between the traditional approaches of quantization before and
after imposing the constraints. Its kinematic aspects uses the
former while its interpretation and dynamical aspects involve the
latter; 4.~As it stands, our method suffers from the
multiple-choice problem and is plagued with the problems related
to the non-Hermiticity of the quantum Hamiltonian for typical
closed universes \cite{review}. In our opinion these problems
cannot be viewed as insurmountable obstacles unless a
comprehensive investigation is performed and a concrete evidence
(such as a no-go theorem) shows otherwise.

\section*{Acknowledgment}

This work has been supported by the Turkish Academy of Sciences in
the framework of the Young Researcher Award Program
(EA-T$\ddot{\rm U}$BA-GEB$\dot{\rm I}$P/2001-1-1).

\end{document}